\documentclass[prl,aps,twocolumn,showpacs,superscriptaddress]{revtex4}
\usepackage{bm}
\usepackage{epsfig}

\usepackage{amsmath}
\usepackage{amssymb}
\usepackage{color}

\newcommand{\ee}{\mathrm{e}}

\begin{document}

\title{Information-theoretic measurements of coupling between structure and dynamics in glass-formers}
\author{Robert L. Jack}
\affiliation{Department of Physics, University of Bath, Bath, BA2 7AY, United Kingdom}
\author{Andrew J. Dunleavy}
\affiliation{HH Wills Physics Laboratory, Tyndall Avenue, Bristol, BS8 1TL, UK}
\affiliation{School of Chemistry, University of Bristol, Cantock Close, Bristol, BS8 1TS, UK}
\affiliation{Bristol Centre for Complexity Sciences, Bristol BS8 1TW, UK}
\author{C. Patrick Royall}
\affiliation{HH Wills Physics Laboratory, Tyndall Avenue, Bristol, BS8 1TL, UK}
\affiliation{School of Chemistry, University of Bristol, Cantock Close, Bristol, BS8 1TS, UK}
\affiliation{Centre for Nanoscience and Quantum Information, Tyndall Avenue, Bristol, BS8 1FD, UK}

\begin{abstract}
We analyse the connections between structure and dynamics in two model glass-formers, using the mutual information between an initial configuration and the ensuing dynamics to compare the predictive value of different structural observables.  We consider the predictive power of normal modes, locally favoured structures, and coarse-grained measurements of local energy and density.  The mutual information allows the influence of the liquid structure on the dynamics to be analysed quantitatively as a function of time, showing that normal modes give the most useful predictions on short time scales while local energy and density are most strongly predictive at long times.
\end{abstract}


\maketitle

As supercooled liquids approach their glass transitions, structural relaxation slows down dramatically, but molecular configurations remain disordered and apparently random~\cite{nagel-review,deb-review}. However, 
 computer simulations~\cite{propensity,coslo07,asaph-modes,malins2013fara,coslo-modes,brito-wyart} 
 and experiments~\cite{candelier2010,leocmach2012} show that liquid structure and dynamical relaxation are correlated in these systems,
  as predicted (or assumed) in several theories~\cite{MCT,tarjus-review,kawasaki-mrco,liu-manning,RFOT-review,bb-ktw-2004,i-mct}. 
 However, correlations between structure and dynamics do not by themselves imply a causal relationship~\cite{ashton-modes}: other
 theories~\cite{GC} assume that local structure plays only a peripheral role in dynamical relaxation.
Correlations between structure and dynamics can be demonstrated at a microscopic 
level~\cite{propensity,coslo07,asaph-modes,malins2013fara,coslo-modes,brito-wyart},
by exploiting the dynamically heterogeneous nature of glassy relaxation~\cite{ediger-dh-review}.
That is, individual particles have different propensities for motion~\cite{propensity},
depending on local structure.  
Here, we use \emph{information theory}~\cite{info-book} to analyze the strength of these correlations, by measuring the extent to which structural measurements can be used to \emph{predict} particle dynamics at subsequent times. This quantitative analysis
provides a stringent test of proposed causal links between structural features and slow dynamics, in contrast to previous analyses based on restricted subsets of particles or snapshots of the system. 
In two model glass-formers, we find that coarse-grained measurements of energy and density~\cite{matharoo06,berthier-predict,berthier-3point} give the most predictive information for long times.  In one of the models, we also find that vibrational modes~\cite{asaph-modes,brito-wyart,coslo-modes,liu-manning} are strongly correlated with motion on relatively short time scales.  Compared to these effects,
the correlation between dynamics and low energy (or low enthalpy) local structures is relatively weak. 

We present results for  the Kob-Andersen (KA) mixture of Lennard-Jones particles~\cite{ka95}, and an equimolar five-component hard sphere (HS) mixture, which  mimics colloidal suspensions~\cite{royall2014}.  
Both systems contain particles of different sizes, with the diameter of the largest particles being $\sigma=1$ (which sets the unit of length). 
The KA system evolves with overdamped (Monte Carlo) dynamics as in~\cite{berthier-mc2007}; we focus on a temperature $T=0.5$. 
The HS system evolves by event-driven molecular dynamics~\cite{dynamo}; we consider volume fractions $\phi$ in the range $0.52-0.58$.  
In both systems, we use $\Delta t$ to indicate the fundamental unit of time.
The relaxation at the state points that we consider is up 
to $3$ decades slower than relaxation at the onset of glassy dynamics, where it is of order $\Delta t$ (in both systems).
Further system details are given in the Supporting Information (SI)~\cite{SI}.

\begin{figure*}
\includegraphics[width=15.8cm]{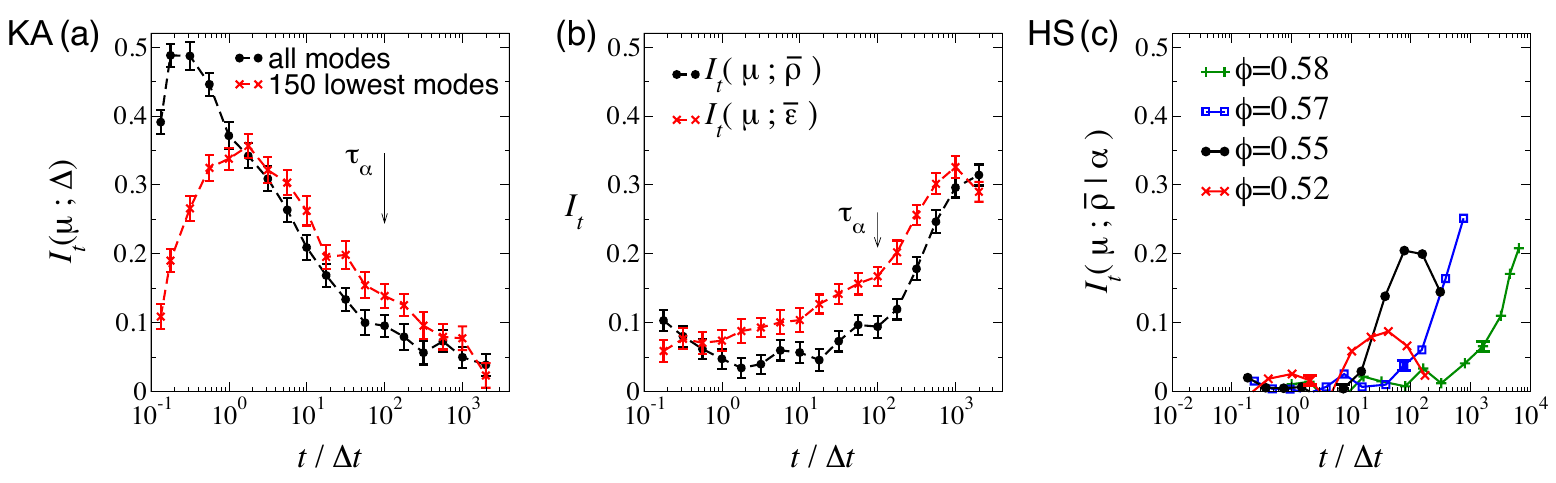}
\caption{
MI measurements $I_t(\mu;s)$ in the KA system and $I_t(\mu;s|\alpha)$ in the HS system.
We show MI between propensity and
(a)~Debye-Waller factors $\Delta_i$ and $\Delta_{i,n=150}$ in the KA system;
(b)~coarse-grained energy and density in the KA system;
(c)~coarse-grained density in the HS system.
The structural relaxation time $\tau_\alpha$ is indicated with arrows in (a,b).  In (c), we show
error bars only for $t=\tau_\alpha$: errors at other times are similar in magnitude.
The behaviour of the MI at long times is discussed in the main text.
}
\label{fig:MI-modes-hydro}
\end{figure*}

To characterize particle dynamics in these systems, we define the dynamical propensity~\cite{propensity} of particle $i$ as $\mu_{i,t} = \langle |\bm{r}_i(t) - \bm{r}_i(t_0)|^2 \rangle_{\rm iso}$, where $\bm{r}_i(t)$ is the particle position at time $t$, and the isoconfigurational average is calculated over many independent dynamical simulations, all with the same initial particle positions but with independent random initial velocities (and independent
stochastic dynamics in the KA system).
The role of the ``lag time'' $t_0$ is discussed in SI~\cite{SI}: we take $t_0\approx0.1\Delta t$. 
We use $s_i$ to denote a structural measurement at time $t=0$, which depends in general on particle $i$ and all particles in its vicinity. 
To quantify the strength of the correlation between $s_i$ and the dynamical propensity $\mu_{i,t}$, we use mutual information (MI) measurements~\cite{info-book}.
The MI is defined
as
\begin{equation}
I_t(\mu;s) =  \sum_s \int\mathrm\!{d}\mu\,p_t(\mu,s) \log_2 \frac{p_t(\mu,s)}{p_t(\mu)p(s)} ,
\label{equ:MI-mu}
\end{equation}
where $p_t(\mu,s)$ is the joint probability distribution of $\mu$ and $s$, while $p_t(\mu)$ and $p(s)$ are its marginal distributions. 
We assumed here that $s_i$ takes discrete values: for continuous attributes $s_i$, the sum over $s$ replaced by an integral.

The MI gives ``the average amount of information about the propensity $\mu_{it}$ that is provided by a measurement of $s_i$''.
Since $s_i$ depends only on the initial condition, the MI measures \emph{predictive} information. 
The MI  may be evaluated
for any structural observable $s_i$, and it makes no assumptions on the nature of the correlation between $\mu_{i,t}$ and $s_i$.  As such, it
represents a generally-applicable figure-of-merit for comparing the influence on dynamics of different structural measures, going beyond previous
comparisons of snapshots~\cite{propensity,coslo-modes,brito-wyart,asaph-modes,berthier-predict,matharoo06} or 
analyses of selected subsets of particles~\cite{coslo07,malins2013fara}.
This use of~(\ref{equ:MI-mu}) as a quantitative measure of information~\cite{info-book} is similar to the use of entropy as a measure of disorder in statistical mechanics, with the role of disorder being taken by the variation in propensity between different particles. Particles with the same value of $s_i$ typically have less variation in their propensity, so specifying $s_i$ reduces the variation in $\mu_{it}$, just as introducing a constraint in statistical mechanics reduces the entropy~\cite{chandler-book}.  The MI is equivalent to this entropy reduction.  Information is conventionally measured in bits, with one bit corresponding to a reduction in entropy of $k_B \ln 2$. Our procedure for estimating MI is described in the SI: the method ensures as far as possible
that we obtain $I_t(\mu;s)=0$ if $\mu$ and $s$ are independent; it also provides an estimate of the numerical uncertainty in the MI.

To illustrate the use of MI, 
let $s_i$ be the type (A or B) of particle $i$ in the KA system.  The different types have different
dynamical relaxation so measuring the particle type provides predictive information about particle dynamics.  In SI~\cite{SI}, we show
that measuring the type of particle $i$ provides between 0.1 and 0.7 bits of information about the propensity $\mu_{i,t}$, 
depending on the time $t$.  This value is a useful baseline in interpreting the results that follow: if a structural measurement is strongly coupled with dynamics, we argue that $I_t(\mu;s)$ should be at least of order 0.1 bit, while MIs much less than this are indicative of weak coupling.

Figure~\ref{fig:MI-modes-hydro} shows MI measurements between particle propensities and several aspects of liquid structure, 
for both KA and HS systems. 
Since the influence of particle type on dynamics is not directly
related to glassy behaviour, we measure mutual information where the predictability based on particle type
has already been taken into account.  That is, for several different $s_i$,
we measure ``the information about $\mu_{i,t}$ that is provided by a measurement of $s_i$, for a particle whose
type is already known''.  In the KA system, we achieve this by restricting the distributions in (\ref{equ:MI-mu}) to particles of type A, which
form the majority ($80\%$) of the system.  In the HS system, we use a `conditional MI', 
$I(\mu ; s | \alpha) = \sum_{s,\alpha} \int\!\mathrm{d}\mu\, p(\mu,s,\alpha) \log_2 \frac{ p(\mu,s|\alpha)}{p(\mu|\alpha) p(s|\alpha) }$
where $\alpha$ indicates the particle type~\cite{SI}.
Our choices of $s_i$ reflect different theoretical pictures of glassy systems: we now discuss the implications of these results for those theories.

Several links have been proposed between normal modes in glassy systems and their material properties~\cite{liu-manning,coslo-modes,asaph-modes,brito-wyart}.  Low-frequency modes in a supercooled liquid define a set of ``soft directions'' on its potential energy surface (or energy landscape), and both thermal fluctuations and structural relaxation couple significantly to these modes~\cite{coslo-modes,asaph-modes,brito-wyart}.  These modes also play a central role in the analogy between glassy behaviour and jamming~\cite{liu-manning}. We analyze them~\cite{SI} by quenching the KA system to its nearest energy minimum (inherent structure), and diagonalizing the Hessian matrix of the energy at that minimum.
The resulting eigenvectors and eigenvalues are $\vec{v}_k$ and $\omega^2_k$, for $k=1 \dots 3N$, and one defines a ``local Debye-Waller (DW) factor''  
$\Delta_i^2 = \sum_k |\bm{v}^i_k|^2/\omega_k^2$ that indicates~\cite{asaph-dw,asaph-modes} 
the expected size of fluctuations in the position of particle $i$, based on an expansion about the energy minimum. (Here $\bm{v}_k^i$ is a vector containing the three components of $\vec{v}_k$ associated with particle $i$.)  Since  low frequency modes couple most strongly to structural relaxation~\cite{asaph-modes}, we also define a generalised DW factor $\Delta_{i,n}^2$, which is calculated using only the $n$ modes with lowest $\omega_k$. In HS systems, normal modes cannot be defined by reference to a potential energy surface so we do not consider them here, although alternative definitions of normal modes are possible~\cite{brito-wyart,liu-manning}.
 
 Figure~\ref{fig:MI-modes-hydro}(a) shows that for relatively short time scales $t\approx \Delta t \ll \tau_\alpha$ in the KA model, the mutual information between propensity and DW factors is large (up to 0.5 bits), 
so $\Delta_{i}^2$ and $\Delta_{i,n=150}^2$ are strongly correlated with particle motion.
This indicates that the normal modes accurately mimic the fluctuations of the system within its initial metastable state. On longer time scales, the information provided by these measurements decreases strongly, but $\Delta_{i,n=150}^2$ still provides more than 0.1 bits at the structural relaxation time $\tau_\alpha$, confirming that the low frequency normal modes do have significant predictive power for structural relaxation~\cite{asaph-modes,brito-wyart,liu-manning}.
 
\newcommand{\rhobar}{\overline{\rho}}
\newcommand{\ebar}{\overline{\varepsilon}}

Coarse-grained energy and density measurements are also correlated with dynamical fluctuations~\cite{berthier-science,berthier-3point,berthier-predict,matharoo06}. We define a local density, coarse-grained on a scale $\ell$, as $\rhobar_i^\ell = \ell^{-3}\sum_j \ee^{-r_{ij}^2/\ell^2}$, where the sum runs over all particles $j$ and $r_{ij}$ is the distance between particles $i$ and $j$~\cite{berthier-predict}.  Similarly, the locally-averaged energy is $\ebar^\ell_i =  \sum_j \varepsilon_j \ee^{-r_{ij}^2/\ell^2}/(\ell^3\rhobar_i^\ell)$ where $\varepsilon_j$ is the energy of particle $j$. Figures~\ref{fig:MI-modes-hydro}(b,c) show that for $\ell=2$ these coarse-grained quantities have strong predictive power on time scales longer than the structural relaxation time, but the MI is smaller for relaxation times up to and including $\tau_\alpha$. The results are broadly similar for both models (for the HS model,
error bars are shown only at $t=\tau_\alpha$, to indicate the $\phi$-dependence of this time scale).  
We show data for $\ell=2\sigma$ since this gives a significant MI throughout this range of data: dependence of the MI on $\ell$ is discussed in SI~\cite{SI}. 

Throughout the glassy regime, we expect $I(\mu;\bar{\rho})$ and $I(\mu;\ebar)$ to have peaks at some time $t^*$,
 before decreasing at longer times (see for example the HS data at $\phi=0.55$).
However, for the largest volume fractions it is clear that $t^*$ is significantly larger than $\tau_\alpha$, and is
larger than our sampling window.  We attribute this large $t^*$ to 
hydrodynamic effects that are largely independent of glassy behavior: regions of size $\ell$ with high density or low energy relax 
on a time scale $\ell^2/D$ where $D$ is a diffusion constant.  One therefore expects relaxation in such regions to
be predictably slower than average up to times $t^* \approx \ell^2/D$, which is significantly larger than $\tau_\alpha$.  
Our focus here is on predictability on time scales of order $\tau_\alpha$, where the system
is has significant dynamical heterogeneity and the motion is complex and co-operative.  For this reason, we have not explored the large-time
hydrodynamic behaviour in detail.  
We do note that for the HS system, the MI at $\tau_\alpha$ increases at large $\phi$, indicating that the coupling of dynamics to local density is increasing as the glass transition is approached, consistent with~\cite{berthier-3point,berthier-science,i-mct}.  However, even for the largest $\phi$, the MI is
less than 0.1 bit at $\tau_\alpha$, although it does grow rapidly for larger times.
 
\begin{figure}
\includegraphics[width=7.5cm]{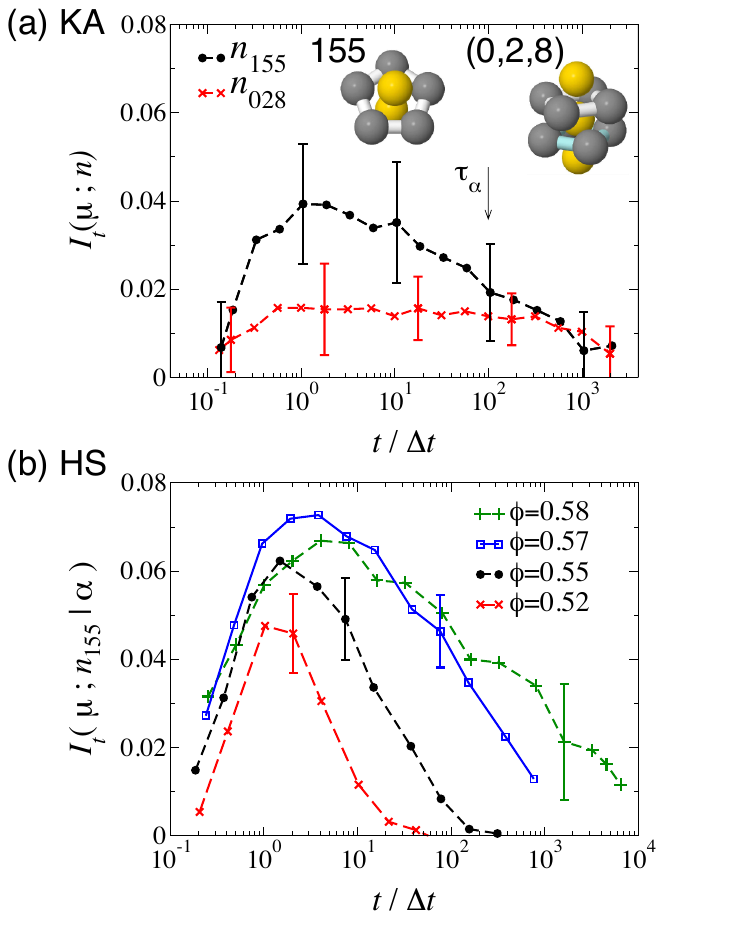}
\caption{MI between propensity and LFS measurements.  We emphasise that
these MIs are smaller in magnitude than those of Fig.~\ref{fig:MI-modes-hydro}, and all are less than 0.1 bit.
(a)~KA system, using LFS with `155' and (0,2,8)  signatures, as described in the main text and illustrated in the figure.
(b)~HS system, using LFS with CNA-155 signatures.  Representative error bars are shown in (a), while in (b) we show error
bars only at $t=\tau_\alpha$.}
\label{fig:MI-lfs}
\end{figure}

An alternative picture of glassy relaxation is based around locally-favoured structures (LFS): atomic or molecular packings that have low energy (or enthalpy)~\cite{tarjus-review}. Particles in LFS typically have slower than average dynamics in glassy 
systems~\cite{coslo07,malins2013fara,malins2013wahn,royall2008}. For both KA and HS models, we consider an LFS based on a pentagonal bipyramid, 
identified by the signature 1551 in the analysis of~\cite{honeycutt}.
These structures are indicative of local fivefold symmetry~\cite{jonsson}. Let $n_{155}(i)$ be the number of pentagonal bipyramids in which particle $i$ participates~\cite{SI}: we expect larger $n_{155}(i)$ to be associated with lower propensity for motion.  In the KA model, we also consider an LFS (bicapped square antiprism) that has been found to be correlated with slow dynamics~\cite{coslo07,malins2013fara}. 
These LFS are associated with Voronoi polyhedra whose signature is $(0,2,8)$ in the notation of~\cite{coslo07}. 
We define $n_{028}(i)=1$ if particle $i$ participates in such an LFS, with $n_{028}(i)=0$ otherwise. 
For the KA model, we calculate $n_{155}$ and $ n_{028}$ using the inherent structure of the system. 

Figure~\ref{fig:MI-lfs}(a) shows results for the KA model, indicating that $n_{155}$ and $n_{028}$ are correlated with particle motion~\cite{coslo07,malins2013fara}.  As with the low-frequency normal modes, the signal is largest on time scales $t\approx \Delta t$ indicative of $\beta$-relaxation, but there is still some correlation at the structural relaxation time.  However, the strength of the correlation is smaller for the LFS than for the normal modes, less than
0.1 bit in all cases. 
Figure~\ref{fig:MI-lfs}(b) shows similar results for the HS system. 
The MI values are larger than those of the KA system, indicating that LFS have  more predictive power for dynamics.  At short times, the MI increases with increasing volume fraction;
however the MI at $\tau_\alpha$ (indicated by the error bars) depends more weakly on $\phi$.
We argue that the small MI values at $\tau_\alpha$ and longer times, and the absence of an increase of the values at $\tau_\alpha$ 
with volume fraction, both indicate that the LFS identified here are more weakly coupled to the dynamics than the normal modes, at least for the degree of supercooling accessed here.

To summarize our findings so far, Figure~\ref{fig:MI-modes-hydro} shows that Debye-Waller factors and coarse-grained measurements of energy and density have significant coupling to dynamics, providing predictive information comparable with measurements of particle type in the KA model.  
However, the information available from the different measurements has very different time-dependences.  
For short times,
the normal mode analysis captures fast vibrational motion accurately, but the predictive power of this analysis decreases strongly with time.
This indicates that
as structural relaxation starts to take place, the `soft directions' for further relaxation quickly diverge from those that were present at $t=0$.
On the other hand, coarse-grained energy and density measurements have almost no predictive value at short times, but the slow decay of 
large-scale hydrodynamic fluctuations means that they can influence particle dynamics quite strongly even on time scales much longer than $\tau_\alpha$:
on these time scales,
almost all memory of the initial structure has been lost, leaving only the hydrodynamic fluctuations in energy or density.
For the state points considered here, Figure~\ref{fig:MI-lfs} shows that LFS measurements have less predictive power for dynamics in these models, 
and that this predictive power is largest on relatively short time scales associated with $\beta$-relaxation.  Our interpretation is that
since the lifetimes of most LFS are less than $\tau_\alpha$~\cite{malins2013fara,malins2013wahn}, the influence of LFS on dynamics
is (in most cases) similarly short-lived, limiting the predictive power of such measurements for dynamics.

\begin{figure}
\includegraphics[width=4.2cm]{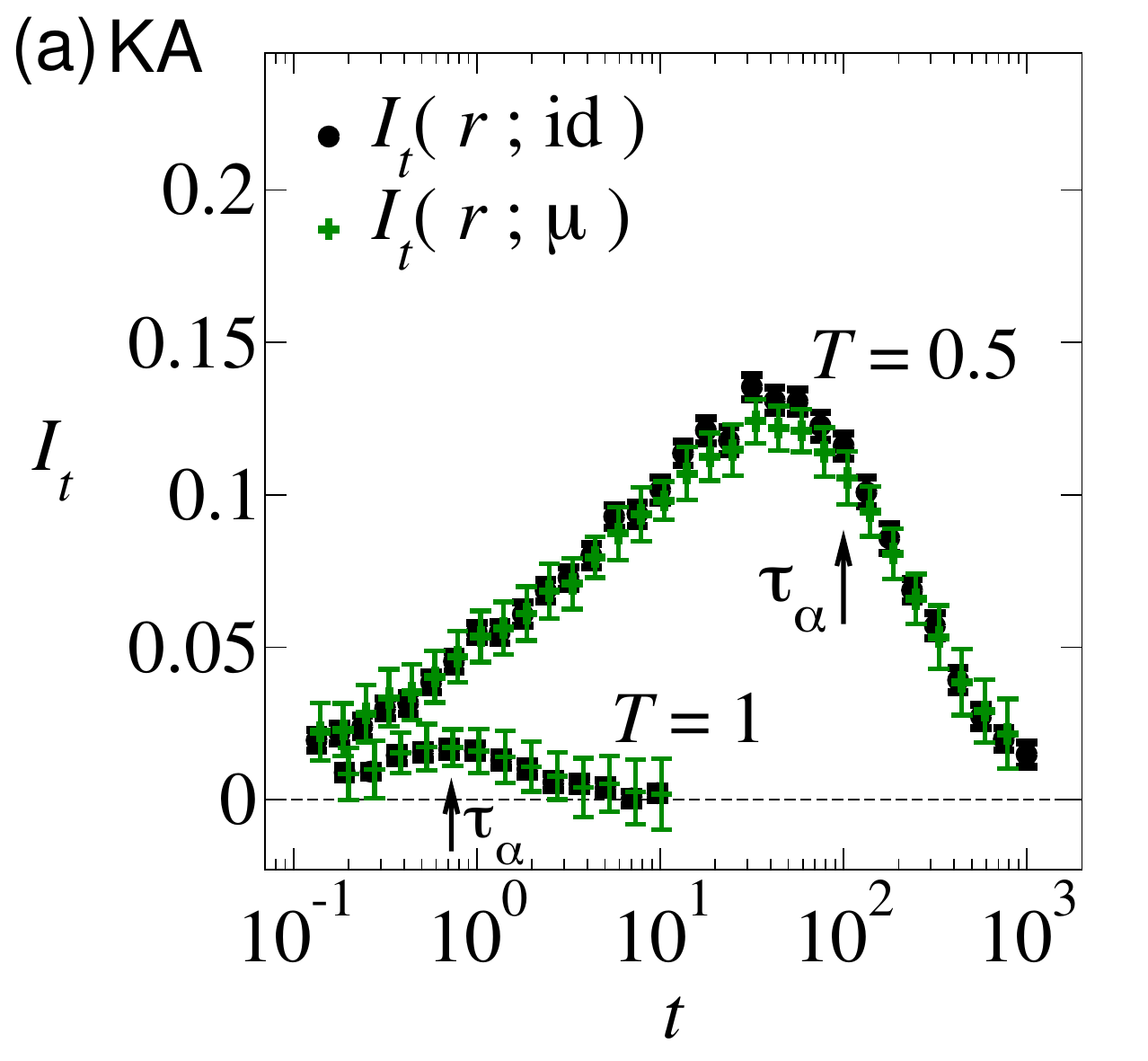}
\includegraphics[width=4.2cm]{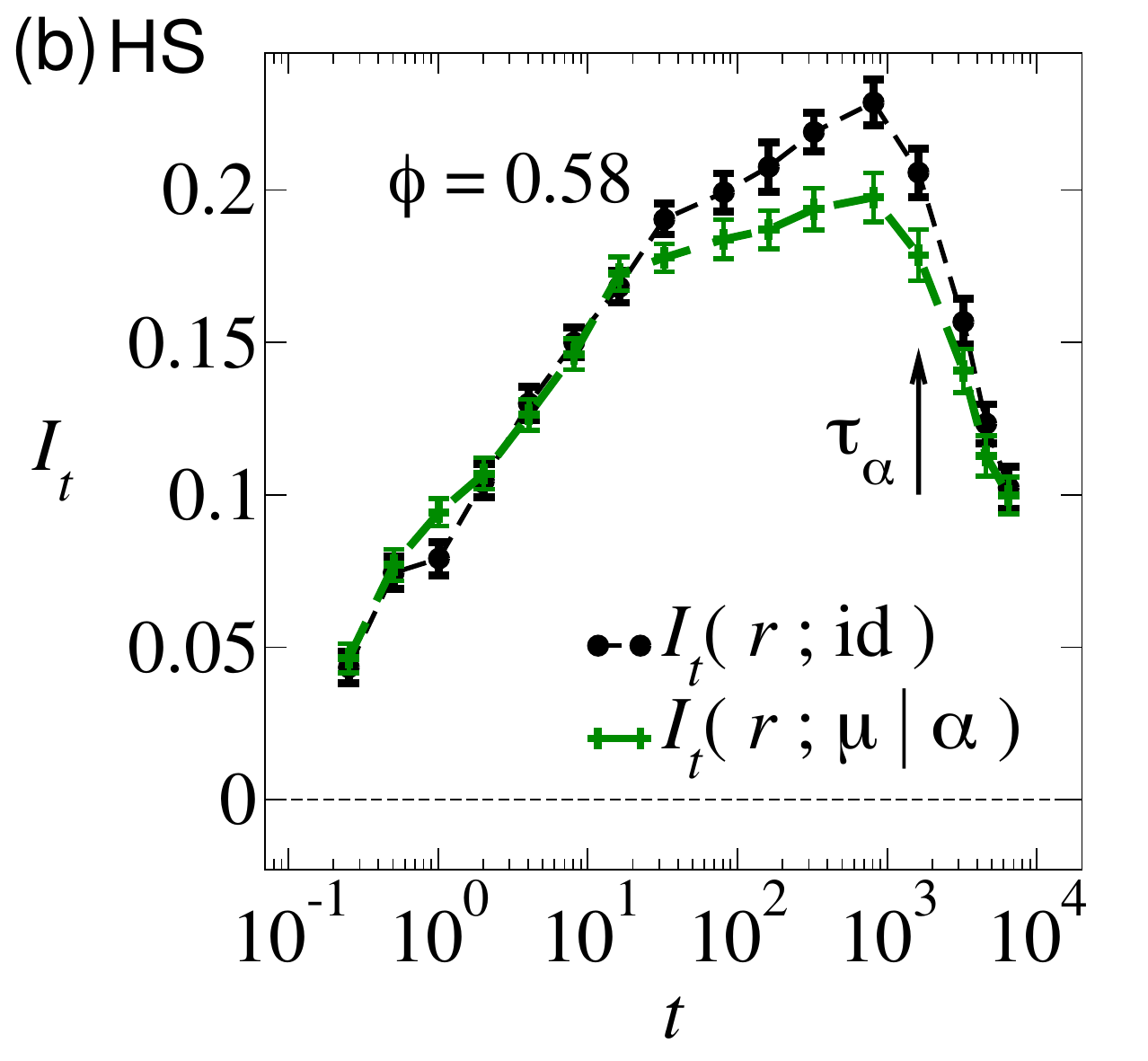}
\caption{Measurements of MI based on particle displacements $r_i$. (a)~KA system for $T=0.5$ and $T=1$; (b)~HS system
at $\phi=0.58$. See text for discussion.
}
\label{fig:MI-dyn}
\end{figure}

Finally, in contrast to the measurements so far, we show how information theory can also be used to analyze how predictable particle motion is in these models, independent of any specific structural observable.
Let $p_{i,t}(r)$ be the (isoconfigurational) distribution of particle displacements $r_i=|\bm{r}_i(t) - \bm{r}_i(t_0)|$. Given data for $N_{\rm p}$ particles (which  may be obtained in general from many initial configurations), 
we define
\begin{equation}
I_t(r;{\rm id}) = N_{\rm p}^{-1} \sum_i \int\!\mathrm{d}r\, p_{i,t}(r) \log \frac{p_{i,t}(r)}{N_{\rm p}^{-1} \sum_i p_{i,t}(r)} ,
\end{equation}
which is the ``average amount of information about a particle's motion that is provided by specifying its initial environment''.  Since a particle's initial environment encodes all predictable aspects of its future motion, $I_t(r;{\rm id})$ indicates how predictable (or reproducible) particle motion is within the system~\cite{propensity,berthier-predict}. Figure~\ref{fig:MI-dyn} shows that $I_t(r;{\rm id})$ is much larger at low temperatures in the KA model than at high temperatures, indicating that structure is more strongly coupled to dynamics at low temperatures.

It is useful to compare $I_t(r;{\rm id})$ with ``the average amount of information about a particle's dynamics that is provided by specifying its propensity'', which is $I_t(r;\mu) = \int\!\mathrm{d}r\,\mathrm{d}\mu\, p_t(r,\mu) \log \frac{p_t(r,\mu)}{p_t(\mu) p_t(r)} ,$ where $p_t(r,\mu)$ is the joint distribution of displacement $r$ and propensity $\mu$, and $p_t(r)$ is the marginal distribution of the displacement.  Since fixing a particle's initial environment necessarily fixes its propensity, one has
\begin{equation}
I_t(r;\mu) \leq I_t(r;{\rm id}) .
\label{equ:mu-i}
\end{equation}
From Fig.~\ref{fig:MI-dyn}, the two quantities in (\ref{equ:mu-i}) are almost equal for the KA model.  Eq.~(\ref{equ:mu-i}) is an ``information-processing inequality''~\cite{info-book}, so this result
 indicates that the propensity captures almost all predictable information about single-particle displacements.  For the HS system, 
 we use a conditional MI between $r$ and $\mu$, to account for particle type, as above.  
The two MIs in (\ref{equ:mu-i}) differ somewhat more strongly than they do in the KA model:
%the propensity $\mu_{i,t}$ captures most of the predictable information about dynamics, but not all of it.
this situation might arise (for example) if some particles have finite average displacements $\langle \bm{r}_i(t) - \bm{r}_i(t_0)\rangle$ that 
are only weakly correlated with their propensities.

Nevertheless, we have $I_t(r;\mu) \approx I_t(r;{\rm id})$ in both models, indicating that
the propensity captures all predictable (reproducible) aspects of the single-particle 
dynamics~\cite{propensity}.  This further validates the use of the mutual information as a general figure-of-merit for evaluating proposed connections between structure and dynamics.  
Given the implications of Figs.~\ref{fig:MI-modes-hydro} and~\ref{fig:MI-lfs} for the strength and time-dependence
of the coupling between structure and dynamics, we hope that future studies will exploit these information-theoretic measurements
to further elucidate which (if any) structural features are responsible for the strong dynamical slowing in supercooled liquids.

We thank Peter Harrowell, Peter Sollich, Gilles Tarjus, and Karoline Wiesner for helpful discussions.  
RLJ and AJD  were supported by the EPSRC through grants EP/I003797/1
and EP/E501214/1 respectively.
CPR gratefully acknowledges the Royal Society for financial support.

\vfill\eject

\begin{appendix}

\section{Supporting Information}

\newcommand{\kacite}{25}
\newcommand{\kaMCcite}{27}
\newcommand{\Coscite}{7}
\newcommand{\TCCcite}{S1}
\newcommand{\CNAcite}{34,35}
\newcommand{\Dynamocite}{28}

\renewcommand{\thefigure}{S\arabic{figure}}
\setcounter{figure}{0}

This supporting information contains:

\begin{itemize}
\item Details of the models described in the main text, and the methods used to identify
locally-favored structures.
\item Illustrative results of mutual information between propensity and particle type
\item Discussion of the $\ell$-dependence of the results shown in Fig.~1(b,c)
\item Analysis of the numerical method that we use when estimating mutual information.
\end{itemize}

\subsection{Model systems}

The KA mixture is defined as in [\kacite].  The system consists of $N=1400$ particles of which $80\%$ are of type
A and $20\%$ of type B.  
The particles interact by Lennard-Jones potentials with parameters $(\epsilon_{\rm AA},\epsilon_{\rm AB},\epsilon_{\rm BB})
=(1.0,1.5,0.5)\epsilon$ and $( \sigma_{\rm AA},\sigma_{\rm AB},\sigma_{\rm BB})
=(1.0,0.8,0.88)\sigma$.  Temperatures are quoted in units of $\epsilon$, with Boltzmann's constant $k_{\rm B}=1$.
The onset temperature of glassy dynamics is $T\approx1$ and the mode-coupling temperature has
been estimated~[\kacite] to be $T=0.435$.
The total number density of particles is $(10/9.4\sigma)^3\approx 1.2\sigma^{-3}$, and we use a
cubic simulation box with periodic boundaries.
The system evolves by Monte Carlo dynamics as in [\kaMCcite], with trial displacements drawn from a cube of side $0.15\sigma$ centred
at the origin.  The mean-square displacement per trial move is $\delta^2 = (0.075\sigma)^2$ and we define the fundamental
time unit $\Delta t=\sigma^2/D_0$ where $D_0$ is the diffusion constant of free particle.  The result is that $\Delta t$
corresponds to $6(\sigma/\delta)^2\approx 1070$ proposed MC moves per particle.  The structural relaxation time $\tau_\alpha$
discussed in the main text is defined as $F_{\rm s}(k,\tau_\alpha)=(1/{\rm e})$ where $F_{\rm s}(k,t)=\langle \ee^{-{\rm i}\bm{k}\cdot[\bm{r}_i(t)-\bm{r}_i(0)]}\rangle$
is evaluated by an average over particles of type A, and $k=|\bm{k}|=7.25/\sigma$.  

The polydisperse hard sphere system consists of an
equimolar mix of five particle species with diameters $(1.000, 0.938,
0.899, 0.861, 0.799)\sigma$, all with equal masses $m$. The particles
interact as hard spheres and the system evolves by
event-driven molecular dynamics (implemented by DynamO~[\Dynamocite]). The system
comprises $N=1372$ particles and the simulation box is cubic with periodic
boundary conditions. The time unit in the system is $\Delta t = \sqrt{m
\sigma^2 / k_{\rm B}T}$.  The structural relaxation time is evaluated at $k=2\pi/\sigma$.

\subsection{Identifying locally favored structures}

Here, we briefly describe the structural measurements $n_{155}$ and $n_{028}$ that we use to identify locally-favoured
structures in these systems.
These measurements are based on Voronoi analyses of the system.  In the
KA model, we perform this analysis after quenching the system to its nearest energy minimum (inherent structure).
We follow [\Coscite] in using a Voronoi analysis where faces between A and B particles are located closer
to the B particles, consistent with their smaller size.  In the HS system, we use a regular Voronoi analysis, in which
faces are midway between neighbouring particles.

We identify $(0,2,8)$ Voronoi polyhedra in the KA system as those with ten faces,
of which exactly two have four edges, and eight have five edges.  This particle and its ten Voronoi neighbours form a cluster
(11A in the topological cluster classification~[\TCCcite]), and we set $n_{028}(i)=1$ for all particles in these clusters.

To identify the pentagonal bipyramids in which particle $i$ participates (in both HS and KA systems), 
we identify $n_{155}(i)$ as the number of pentagonal faces on the Voronoi cell of that particle.  This gives the number
of neighbours of particle $i$ that share exactly five mutual neighbours with particle $i$.  The procedure is equivalent to counting
the number of `1551' bonds in the common neighbour analysis (CNA)~[\CNAcite], and is similar to the identification of `7A' clusters
in the topological cluster classification~[\TCCcite].

\subsection{MI between particle type and propensity}

\begin{figure}
\includegraphics[width=7cm]{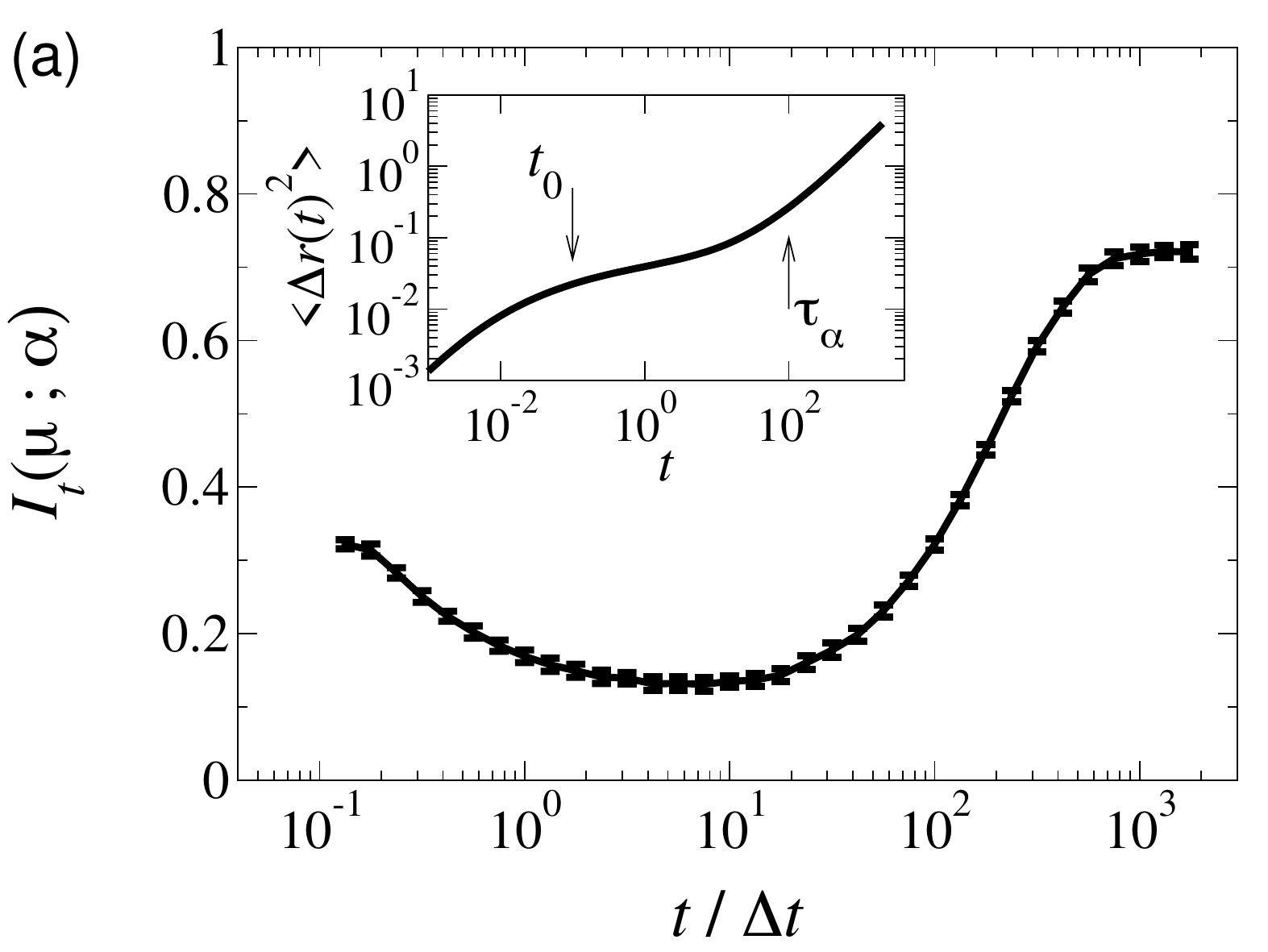}
\includegraphics[width=7cm]{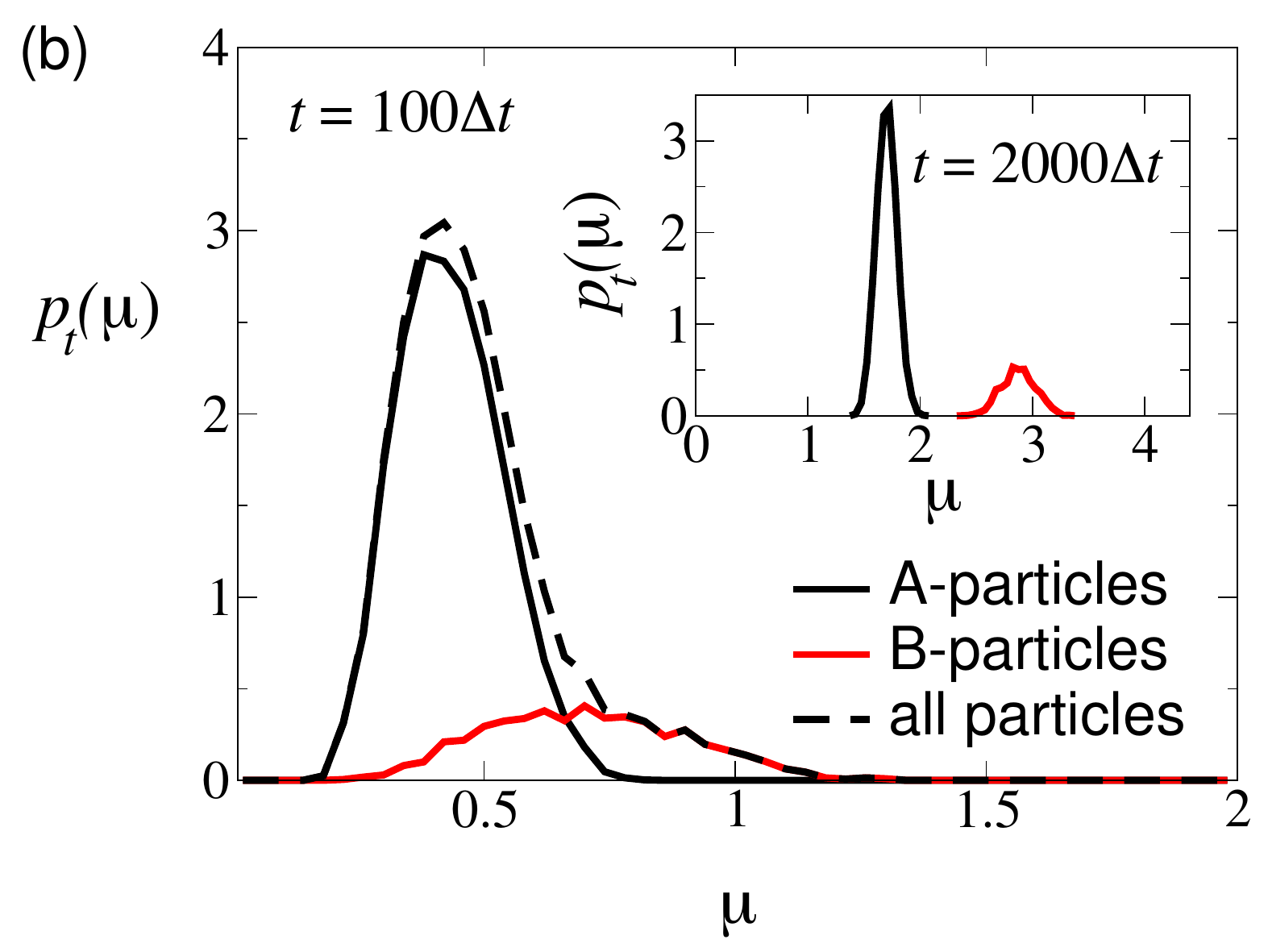}
\caption{Illustration of MI measurements.
(a)~Mutual information between the propensity $\mu$ and the particle type $\alpha$ in the KA system. (Inset)~Mean square displacement of all particles, with arrows indicating the structural relaxation time $\tau_\alpha\approx 100\Delta t$ and the lag time
$t_0=0.1\Delta t$. (b)~Distributions of the propensity at $t=100\Delta t$ and $t=2000\Delta t$ (inset): see text for discussion.
%At the earlier time, the distributions of the propensity for the two types of particle are overlapping while for long times there is almost no overlap, resulting in a saturation of the mutual information to $I_{1/5}\approx 0.72$ bits (see main text).
}
\label{fig:MI-type}
\end{figure}

To demonstrate the physical meaning of MI, Fig.~\ref{fig:MI-type}(a) shows $I_t(\mu;s)$ for the KA system, where the structural measurement $s_i$ is taken to be the particle type $\alpha_i={\rm A,B}$.  The B-particles are more mobile in this system, and as shown in the inset, at large times $t\gg\tau_\alpha$, the propensity distributions for the two kinds of particle have almost zero overlap.  Thus, for these very long times, measuring the particle type splits the propensity distribution into two distinct components: this provides $-f \log_2 f - (1-f)\log_2 (1-f)$ bits of information (similar to a mixing entropy), where $f$ and $(1-f)$ are the fractions of particles in each component. If the components were equal in size, the MI would be exactly 1 bit: here the B particles are less numerous ($f=0.2$) so the MI is less, approximately $0.7$ bits.
For times $t$ close to the structural relaxation time $\tau_\alpha$, Fig.~\ref{fig:MI-type}(b) shows that the propensity distributions of the two types differ from each other, but there is a region of significant overlap. In this case, measuring the particle type provides $0.3$ bits of information about $\mu_{it}$.

\subsection{Predictive power of local energy/density: dependence on length scale $\ell$}

\begin{figure}
\includegraphics[width=6cm]{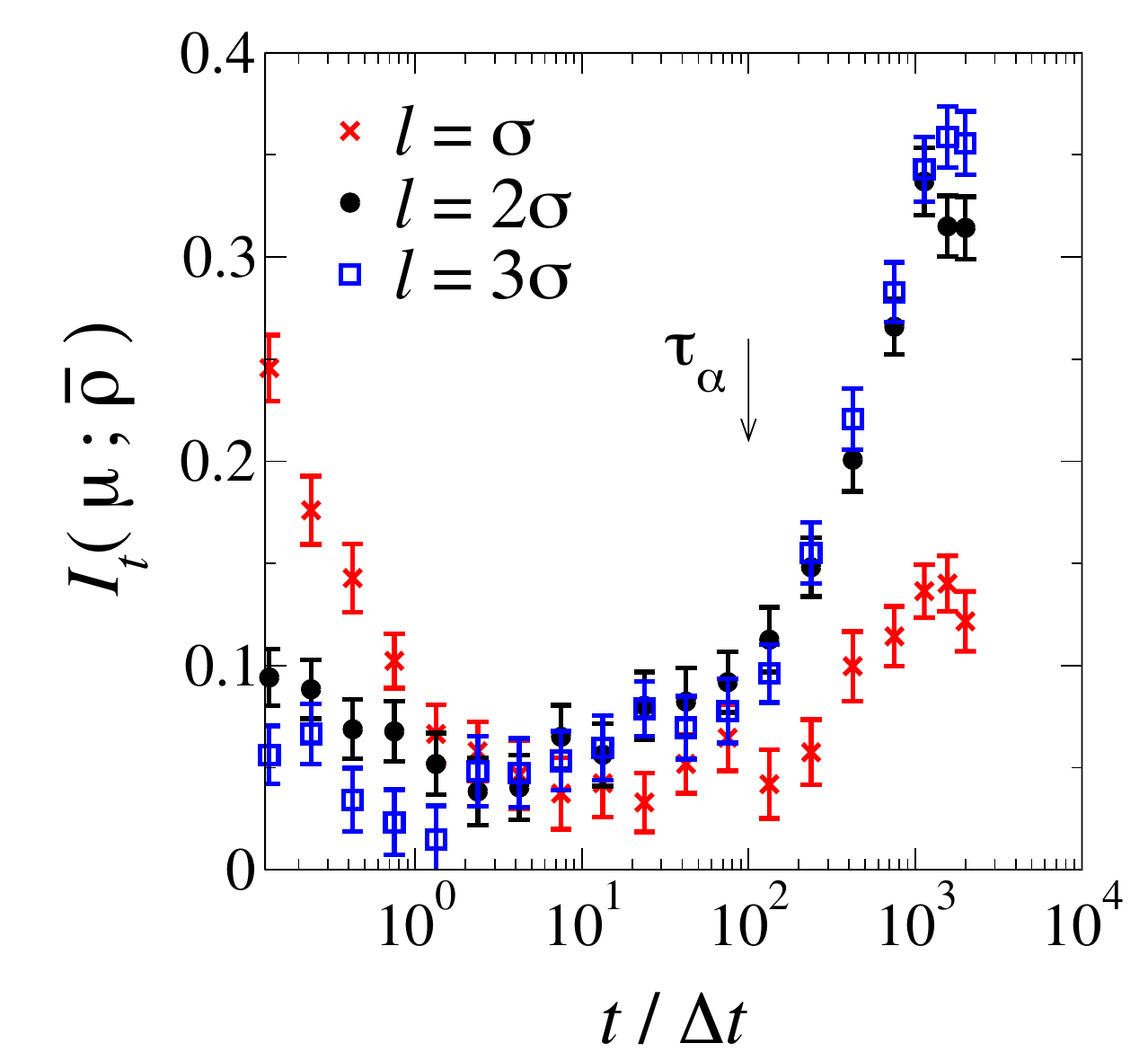}
\caption{MI between coarse-grained density $\rhobar^\ell$ and propensity in the KA model at $T=0.5$, 
for $\ell$ between $\sigma$ and $3\sigma$.  [The data for $\ell=2\sigma$ are also shown in Fig. 2(c) of the main text.]
See text for discussion.}
\label{fig:supp-mi-ell}
\end{figure}

The coarse-grained measurements of local energy and density $\rhobar$ and $\ebar$ discussed in the main text
depend (by definition) on a length scale $\ell$
that indicates the size of the local coarse-graining region.  The effect of varying this length scale is illustrated in Figure~\ref{fig:supp-mi-ell},
where the MI between propensity and $\rhobar^\ell$ is plotted, for various $\ell$.
For short lengths (here, $\ell=\sigma$), the local density is strongly predictive on short times, presumably due to the influence
of free volume on the vibrational motion of particles within a single metastable state.  For longer times, the
MI increases with $\ell$ before saturating at $\ell\approx 2\sigma$.  
In the long time regime, and for all cases considered, the MI for $\ell=2\sigma$ is almost always larger than
for smaller $\ell$-values, and increasing $\ell$ above $2\sigma$ does not significantly increase the MI (as in Fig.~\ref{fig:supp-mi-ell}).
This observation motivated our choice of $\ell=2\sigma$ for the data shown in Figs.~2(c,f).
The MI between coarse-grained energy and propensity does not show the early-time signal found for $I(\mu;\rhobar)$ at $\ell=\sigma$ but
otherwise behaves similarly to $I(\mu;\rhobar)$.

\subsection{Estimating mutual information}

\newcommand{\SEntLast}{S6}

Calculating mutual information (MI) from numerical data requires some care, since estimators are vulnerable
to systematic errors if sample sizes are not sufficiently large.  A variety of estimators have been developed (see for example~[S1-\SEntLast]),
many of which use Bayesian methods, exploiting prior knowledge (or assumptions) about the form of the underlying distributions 
in order to better estimate either entropies or mutual informations~\cite{nsb,kennel,pillow13}.  
In this work, we use a simple method that we have tailored to the problem of
interest here, based on the method of~\cite{kraskov}.

In all measurements, we discretise the propensity, forming a histogram with bins of width $\delta\mu=\langle \mu\rangle/10$,
where $\langle \mu\rangle$ is the mean propensity.  The width of the bins is comparable with the numerical
uncertainties in our estimates of the $\mu_i$, which are obtained from between 100 and 250 independent trajectories.
For this reason, storing the propensities to greater accuracy than the bin width would not make our measurements of MI any more accurate 
-- the binning does not 
introduce numerical artefacts, and is convenient in what follows.  Further, since the same binning is used for all MI measurements,
we are able to make a fair comparison between the different structural measurements shown in Figures 1 and 2 of main text.
In the following, we use $m_i$ as an integer-valued label for the bin in which the propensity $\mu_i$ is located (for example, one may take 
$m_i=\lfloor\mu_i/\delta\mu\rfloor$, 
 the largest integer that is less than or equal to
$\mu_i/\delta\mu$). 

\subsubsection{Two discrete variables}

We first describe the estimator that we use for calculating MI between two discrete-valued variables.  We have in mind that $s_i$ is a
structural observable with a discrete set of possible values, while $m_i$ is the propensity
bin-index as described above.  However, the discussion is general for joint distributions of discrete random variables. 
For each particle, suppose that
we measure two integers $m_i$ and $s_i$.   Then given data for $N_{\rm p}$ particles, let $n(m,s)$ be the number
of particles with $(m_i,s_i)=(m,s)$; also let $n(m)=\sum_s n(m,s)$ be the number of particles with $m_i=m$,
and similarly $n(s)=\sum_m n(m,s)$.  The simplest MI estimate based on these data is the ``plugin estimator'':
\begin{equation}
I^0 = N_{\rm p}^{-1} \sum_{m,s} n(m,s) \log_2 \frac{n(m,s) N_{\rm p}}{n(m)n(s)}
\end{equation}
where the sum runs over all pairs $(m,s)$ for which $n(m,s)>0$.  Given sufficient data,
$I^0$ converges to the mutual information $I(m;s)$, however, this convergence is often quite slow, requiring very 
large $N_{\rm p}$ for an accurate estimate.
In particular, even if the data set is constructed so that $m_i$ and $s_i$ are independent,
one typically finds $I^0>0$, recovering $I^0\to0$ only as $N_{\rm p}\to\infty$.

To see the reason for this, it is useful to write
$I^0 = H^0(m) - H^0(m|s)$ 
with 
\begin{equation}
H^0(m|s) = - \sum_s \frac{n(s)}{N_{\rm p}} \sum_m n(m|s) \log_2 n(m|s)
\end{equation}
where $n(m|s) = n(m,s)/n(s)$,
and 
\begin{equation}
H^0(m) = - \sum_{m}  \frac{n(m)}{N_{\rm p}} \log_2 \frac{n(m)}{N_{\rm p}}
\end{equation}

The key point is that for large enough data sets ($N_{\rm p}\to\infty$) one has $n(m)/N_{\rm p}\to p(m)$ by the law of large numbers,
so that
$H^0(m)$ is an entropy estimator for $H(m) = -\sum_m p(m) \log_2 p(m)$.  Similarly,
$H^0(m|s)$ converges to a weighted sum of conditional entropies of the form $\sum_m p(m|s) \log_2 p(m|s)$, as long as
$n(s)\to\infty$ for all $s$.
The difficulty is that the convergence of 
$H^0(m)$ and $H^0(m|s)$ to their respective limits are ruled by different large parameters ($N_{\rm p}$ and $n(s)$), and there are systematic
errors associated with this convergence if these parameters are not large enough.
In general, $H^0(m)$ and $H^0(m|s)$ both underestimate the relevant entropies, but the
error on $H^0(m)$ is smaller, resulting in a positive systematic error for $I^0(m;s)$.

\begin{figure}
\includegraphics[width=6cm]{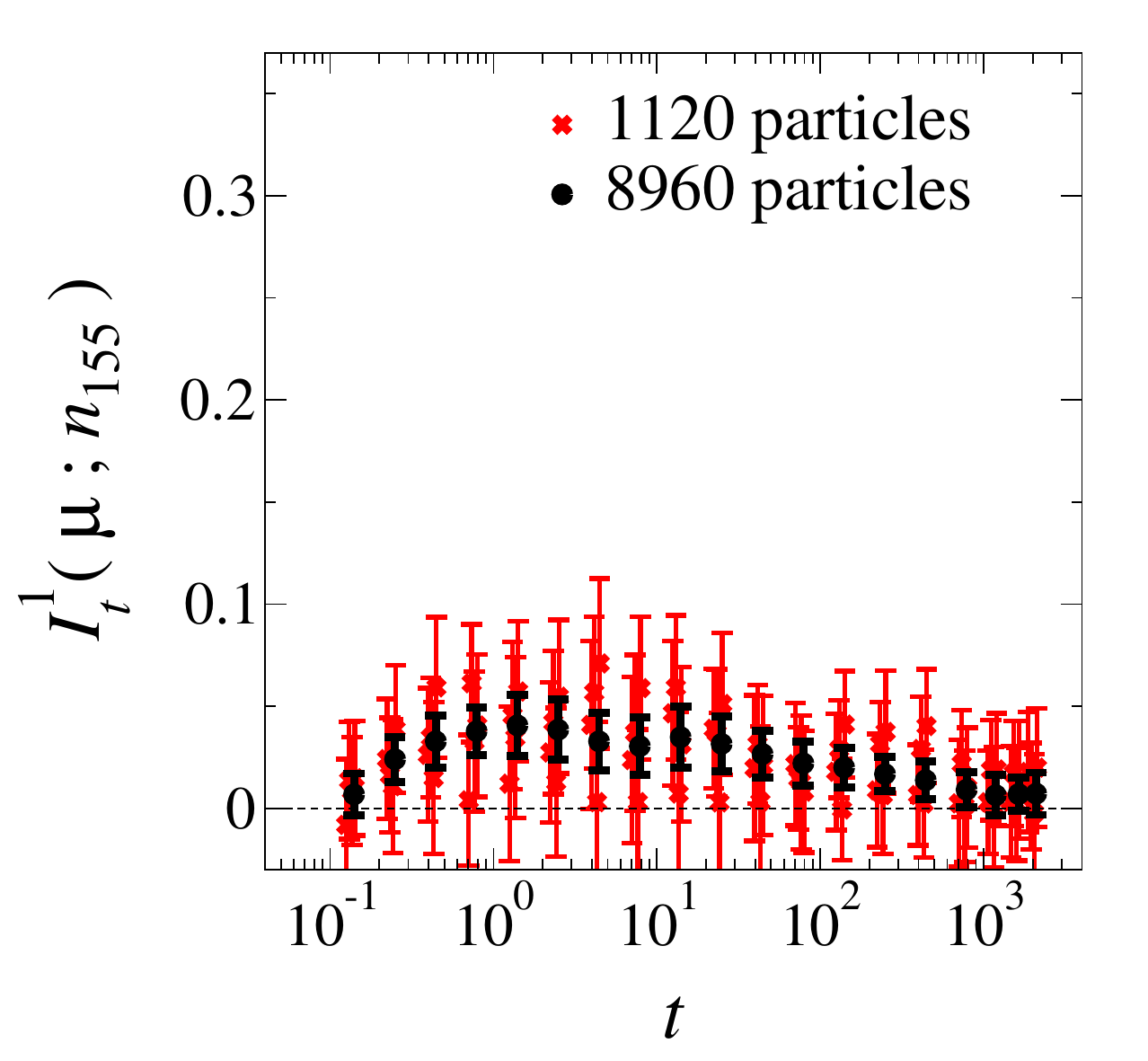}
\includegraphics[width=6cm]{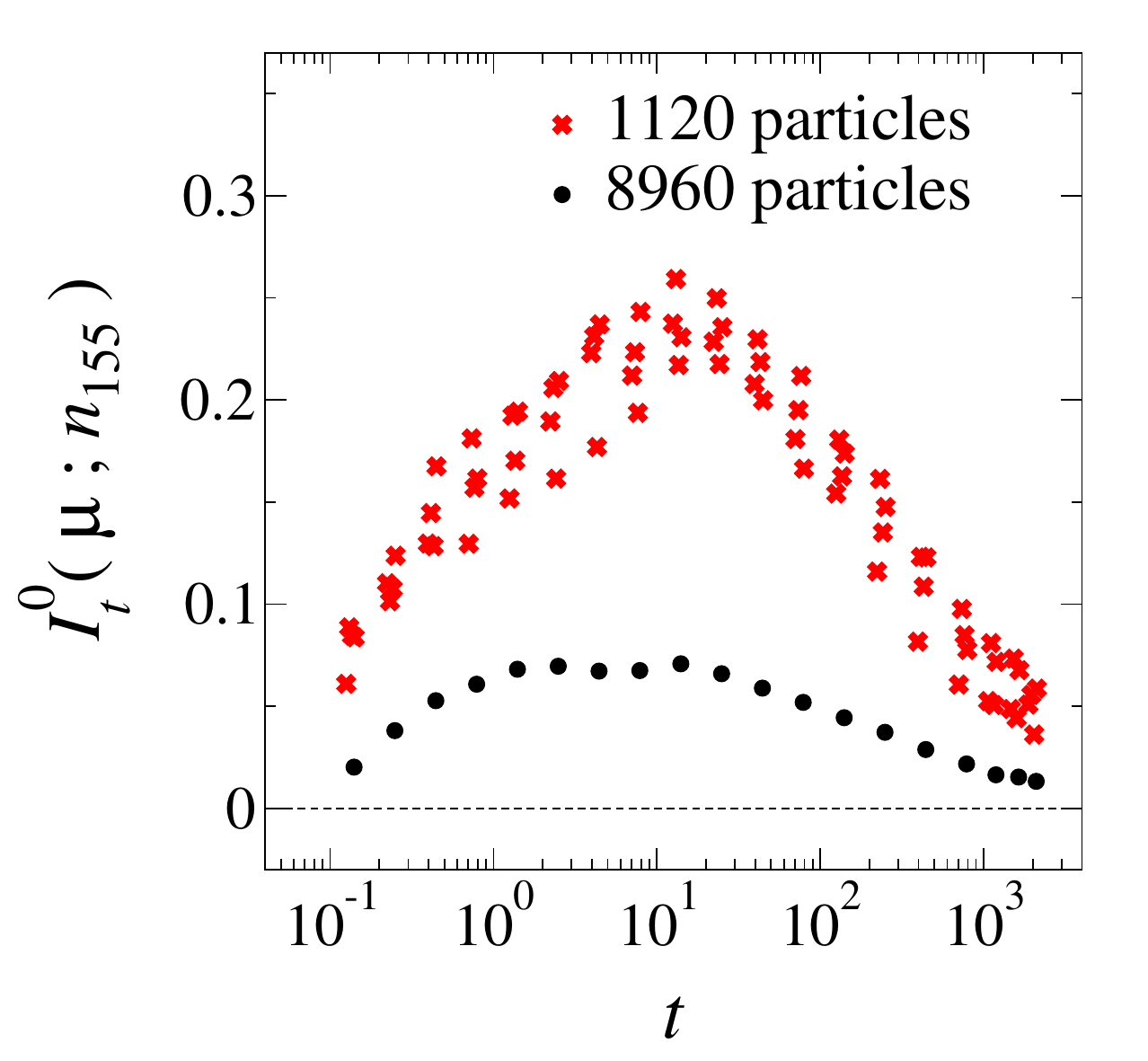}
\caption{(a)~Estimate of MI between $n_{155}$ and the propensity, using the estimator $I^1$.  We show one estimate based on
a sample of $(8\times 1120)=8960$ type-A particles (from 8 independent initial conditions), and four estimates
each based on $1120$ particles (each estimate based on just one initial condition).  The error bars are conservative
estimates of numerical uncertainty (see text) and are reduced on including
more data. The mean estimate has only a weak drift with sample size indicating that systematic errors arising from small 
samples are small.  (b)~Estimates of the same MI based on the same data but using
the estimator $I^0$.  The estimation is poor in this case: 
there is a strong systematic error coming from the finite sample size. 
The error decreases as more data is included but the largest sample
used is not large enough to saturate the limit $N_{\rm p}\to\infty$
and does not provide a reliable estimate of the MI.  
}\label{fig:cna-estim}
\end{figure}

To reduce this effect, we define an alternative estimator
\begin{equation}
I^1 = \sum_s \frac{n(s)}{N_{\rm p}} \sum_{m} n(m|s) \left[\log_2 n(m|s)  - \log_2 \frac{\tilde{n}_s(m)}{n(s)}\right]
\label{equ:i1}
\end{equation}
Here, the $\tilde{n}_s(m)$ are obtained from a random
null data set, as follows.  For each $s$, we draw a random sample of size $n(s)$ (without replacement) from the original
data set, and $\tilde{n}_s(m)$ is defined as the number of particles in that random sample that have $m_i=m$.  For a large enough sample,
$\tilde{n}_s(m)/n(s) \to n(m)/N_{\rm p} \to p(m)$.  However, the advantage of the method is that the convergence of $n(m|s)$ to $p(m|s)$
and $\tilde{n}_s(m)/{n(s)}$ to $p(m)$ are now ruled by the same parameter $n(s)$, and the result is that the systematic errors arising from  the two
terms in (\ref{equ:i1}) tend to cancel each other.  
We note that if instead of drawing a random sample we simply set $\tilde{n}_s(m)/n(s)=n(m)/N_{\rm p}$, independent of $s$, then we recover 
the original estimator $I^0$.  
It is also notable that if $m$ and $s$ are independent then the conditional distribution $n(m|s)$ should be statistically equivalent to the
distribution obtained in the random sample $\tilde{n}_s(m)$.  This means that $I^1$ is free from systematic error in the case where $s$ and $m$
are independent.  This is the most important case for the calculations of this paper, because the MI values found are typically quite small,
and systematic errors when $I\approx 0$ correspond to false-positive signals of correlation between structure and dynamics, which can be misleading.

For each estimate of MI, we compute $I^1$ using several random null data sets (typically 100 realisations are sufficient).
The average value of $I^1$ over the realisations provides our estimate of $I$ while
the standard deviation among the values of $I^1$ gives an estimate on the uncertainty of this estimate.  We
therefore use this standard deviation as the error bar for the estimate of $I$.  We emphasise that the $n(m|s)$ are determined by
the original data and are the same for every realisation of the null data: it is 
the finite size of this original data set
that introduces a finite uncertainty on estimates of $I$. This uncertainty is not reduced by repeated 
sampling over different null data sets, so it is the standard deviation of $I^1$ that gives the relevant error estimate, not the
standard error.

Fig.~\ref{fig:cna-estim} shows estimates for the MI between $n_{155}$ and the propensity, obtained by the estimators $I^1$ and $I^0$, for data
sets of two different sizes.  It can
be seen that $I^0$ is not sufficient for the purposes used here, even for the larger data set,
while $I^1$ gives a consistent estimate of the MI for data sets of both sizes 
considered. The estimate of the uncertainty based on $I^1$ is also self-consistent, in that error bars from independent estimates typically
overlap with each other.

An alternative to (\ref{equ:i1}) can be obtained by interchanging $s$ and $m$, since the MI is symmetric:
\begin{equation}
I'^1 = \sum_m \frac{n(m)}{N_{\rm p}} \sum_{s} n(s|m) \left[\log_2 n(s|m)  - \log_2 \frac{\tilde{n}_m(s)}{n(m)}\right]
\label{equ:i1-prime}
\end{equation}
We use $I^1$ for all calculations of MI between discrete variables, but it is useful to define $I'^1$ in preparation for later sections.  
%(I think $I^1$ is best because it's nice if all $n(s)$ are large.)

\begin{figure}
\includegraphics[width=6cm]{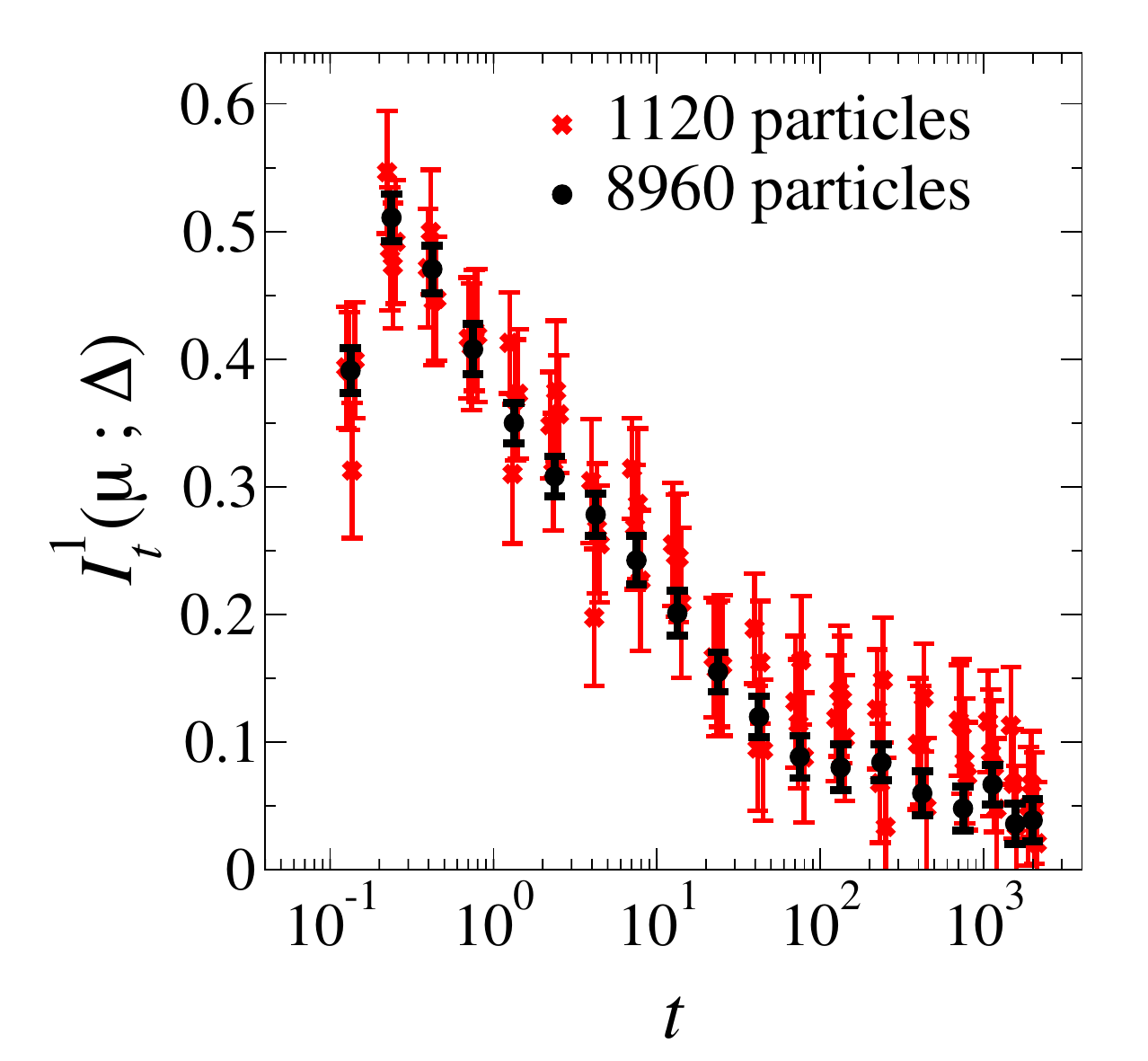}
\caption{Estimate of MI between $\Delta^2_i$ and the propensity, using the estimator $I^2$.  We show one estimate based on
a sample of $(8\times 1120)=8960$ type-A particles (from 8 independent initial conditions), and four estimates
each based on $1120$ particles (from independent initial conditions).  The error bars are conservative
estimates of numerical uncertainty (see text) and are reduced on including
more data. Systematic errors from small samples are weak.
}\label{fig:dw-estim}
\end{figure}

\subsubsection{One discrete and one continuous variable}

We now turn to the case where the structural variable of interest takes continuous values.  The analogue of the estimator $I'^1$ in (\ref{equ:i1-prime}) is
\begin{equation}
I^2 = \sum_m \frac{n(m)}{N_{\rm p}} \, F_{s|m}
\end{equation}
where $F_{s|m}$ is an estimator for $\int\mathrm{d}s \, [p(s|m) \log_2 p(s|m) - p(s) \log_2 p(s)]$.  To define $F_{s|m}$, 
we again take a random sample of size $n(m)$ from the original data,
to ensure that systematic errors on estimates for $\log p(s|m)$ and $\log p(s)$ should cancel as far as possible.
Then, if $s^1,s^2,\dots$ is an ordered list of the values of $s_i$ for those particles with $m_i=m$, and $\tilde{s}^1,\tilde{s}^2,\dots$
is an ordered list of the values of $s_i$ in the random sample, we define
\begin{equation}
F_{s|m} = \frac{1}{[n(m)-1]\ln2}\sum_{i=1}^{n(m)-1} [ \ln(s^{i+1}-s^i) - \ln(\tilde{s}^{i+1}-\tilde{s}^i) ].
\end{equation}
This converges to the required result as $n\to\infty$ because if $x^1,x^2,\dots x^n$ is an ordered list of independent random samples from $p(x)$ then, 
as $n\to\infty$,
one has
\begin{equation}
 \frac{1}{n-1}\sum_{i=1}^{n-1} \ln(x^{i+1}-x^i) +\psi(1) - \psi(n) \to -\int\mathrm{d}x\, p(x)\ln p(x) 
\end{equation}
where $\psi(n)$ is the digamma function, which satisfies $\psi(n+1)-\psi(n)=1/n$ and
$\psi(1)=-\gamma$ where $\gamma=0.577\dots$ is Euler's constant~\cite{kraskov}. 

Fig.~\ref{fig:dw-estim} shows results using the estimator $I^2$.  As with $I^1$, the uncertainty on the estimate of MI is reduced
on including more data, and the systematic variation on increasing $N_{\rm p}$ is weak, indicating that the estimator is reliable.

\end{appendix}

\end{document}